\documentstyle[spie]{article} 
\input{psfig}   

\title{Monitoring the Health and Safety of the ACIS Instrument
On-Board the Chandra X-ray Observatory} 

\author{Shanil N. Virani\supit{a}, Peter G. Ford\supit{b}, Joseph M. DePasquale\supit{a}, and  Paul P. Plucinsky\supit{a}
\skiplinehalf 
\supit{a}Harvard-Smithsonian Center for Astrophysics,
Cambridge, MA  02138
\\
\supit{b}Center for Space Research, Massachusetts Institute of Technology,
Cambridge, MA  02138
}

\authorinfo{Further author information: (Send correspondence to SNV) \\
SNV: E-mail: svirani@cfa.harvard.edu \\
JDP: E-mail: depasq@head-cfa.harvard.edu \\
PPP: E-mail: pplucinsky@cfa.harvard.edu \\
PGF: E-mail: pgf@space.mit.edu \\
}

\begin{document} 
  \maketitle 

\begin{abstract}
	The \textit{Chandra X-ray Observatory} (CXO), NASA's latest
``Great Observatory'', was launched on July 23, 1999 and reached its 
final orbit on August 7, 1999. The CXO is in a highly elliptical
orbit, approximately 140,000 km x 10,000 km, and has a period of 
approximately 63.5 hours ($\approx$ 2.65 days). Communication with the
CXO nominally consists of 1-hour contacts spaced 8-hours apart. Thus,
once a communication link has been established, it is very important
that the health and safety status of the scientific instruments as
well as the Observatory itself be determined as quickly as possible.

In this paper, we focus exclusively on the automated health and safety
monitoring scripts developed for the Advanced CCD Imaging Spectrometer
(ACIS) to use during those 1-hour contacts. ACIS is one of the two focal
plane instruments on-board the CXO. We present an overview of the
real-time ACIS Engineering Data Web Page and the alert schemes
developed for monitoring the instrument status during each
communication contact. A suite of HTML and PERL scripts monitors the
instrument hardware house-keeping electronics (\textit{i.e.}, voltages and currents) and
temperatures during each contact. If a particular instrument component
is performing either above or below pre-established operating
parameters, a sequence of email and alert pages are spawned to the
Science Operations Team of the Chandra X-ray Observatory Center so
that the anomaly can be quickly investigated and corrective actions
taken if necessary. We also briefly discuss the tools used to monitor
the real-time science telemetry reported by the ACIS flight software.

The authors acknowledge support for this research from NASA contract 
NAS8-39073.

\end{abstract}



\section{INTRODUCTION}
\label{sect:intro}  

Just past midnight on July 23, 1999, the space shuttle
\textit{Columbia} lifted-off from Cape Canaveral, Florida. In its payload
bay lay the \textit{Chandra X-ray Observatory} (CXO), the primary
cargo of the \textit{STS-93} mission. Just under 8 hours after 
launch, \textit{Chandra} was
deployed from the space shuttle. However, it would be nearly two
weeks later, after an Inertial Upper Stage booster ``burn'' and several
``burns'' by its own propulsion system, that \textit{Chandra} would reach its 
final orbit. The CXO is now the third of NASA's ``great
observatories'' in space.

The CXO's operational orbit has an apogee of approximately 140,000 km
and a perigee of nearly 10,000 km, with a $28.5^\circ$ initial inclination.
The CXO's highly elliptical orbit, with an orbital period of
approximately 2.65 days, results in high observing efficiency.
Moreover, the fraction of the sky occulted by the Earth is
small over most of the orbital period, as is the fraction of time when
the detector backgrounds are high as the CXO dips into the
Earth's radiation belts. Consequently, approximately 85\% of \textit{Chandra}'s
orbit is available for observing. In fact, uninterrupted observations
lasting as long as 2.3 days are possible\cite{sodell98}.

The CXO carries two focal plane science instruments: the High
Resolution Camera (HRC) and the Advanced CCD Imaging Spectrometer
(ACIS). The Observatory also possesses two objective transmission
gratings: a Low Energy Transmission Grating (LETG) that is  
primarily used with the HRC, and the High Energy Transmission Grating 
(HETG) that is primarily used with the ACIS. In addition to
these instruments,  \textit{Chandra} also carries a radiation monitor -- the
\textit{Electron, Proton, Helium Instrument} (EPHIN) particle detector.

In order to maintain this high level of observing efficiency, a robust
monitoring and alert system must be in place to verify the health and
safety status of not only the instruments themselves, but the whole
observatory as well. This requirement is a result of the fact that the
CXO is not in constant communication with the ground. Communication
with the CXO is obtained through NASA's Deep Space Network (DSN) of 
ground-tracking stations. Since the DSN supports satellite operations of many
different NASA missions, this resource is shared amongst all
satellites currently in space.

\textit{Chandra}'s nominal communication schedule consists of 1-hour
contacts spaced approximately 8-hours apart. Therefore, over the
course of one 24-hour period, the \textit{Chandra} X-ray Observatory Center
(CXC) receives telemetry from the observatory three times per day for
a total of three hours, on average. 
Of course, during the periods in which the ground is not in
contact with \textit{Chandra}, routine science operations continues
autonomously on-board the spacecraft. Each week, a new set of command
loads is generated, reviewed, and uplinked to \textit{Chandra} for use in the
following week. These command loads contain all the necessary
information for the observatory to execute and perform a week's worth
of science observations and spacecraft maintenance activities. In a
companion paper\cite{depasq}, we review the command load inspection
procedure for the ACIS instrument. The science data and observatory
health and safety telemetry are stored on \textit{Chandra}'s solid state
recorders (SSR). During a real-time communication contact, these data are
telemetered to the ground and transfered to the CXC for data
processing (via the Jet Propulsion Laboratory). 
Another paper in this volume\cite{spitzbart} discusses the
monitoring and trends analysis of the SSR data. However, since it can
take as much as 12-24 hours for the SSR data to be processed, analysed, and
hence a putative anomaly found, it is vital that the health and safety
status of the Observatory be quickly established during each
communication pass. In addition to telemetering SSR data, ``live'' or
real-time telemetry data from the observatory and all instruments, 
\textit{i.e.}, the current values of all the
observatory and instrument house-keeping electronics, are sent 
immediately to the \textit{Chandra} Operations and Control Center (OCC)
during each DSN communication pass.

In this paper, we focus exclusively on the real-time monitoring of the Advanced
CCD Imaging Spectrometer (ACIS) on-board the CXO during these one hour
communication contacts. In Section 2, we provide a brief synopsis of
the CXO and its primary focal plane instruments while Section 3 illustrates
the data flow from spacecraft to monitoring. In Section 4 we discuss
the method by which we arrived at our set of health and safety
limits. In Section 5, we present the real-time monitoring
schemes developed for the ACIS instrument. Our conclusions follow in
Section 6.

\section{CHANDRA X-RAY OBSERVATORY'S FOCAL PLANE INSTRUMENTS} 
\label{chandra}

The observatory consists of a spacecraft system and a
telescope/science-instrument payload. The spacecraft system provides
mechanical controls, thermal control, electric power,
communication/command/data management, and pointing and aspect
determination. This section briefly describes the two
focal plane instruments on-board the CXO, the HRC and the ACIS, while
the main emphasis in this paper is the real-time health and safety 
monitoring of the ACIS instrument. The
\textit{AXAF Observatory Guide}\cite{obsguide} and the 
\textit{AXAF Science Instrument Notebook}\cite{SIN} contain a wealth
of information about the \textit{CXO} and its scientific instruments. 
More in depth discussions of the Chandra mission, spacecraft, 
other instruments and subsystems are presented elsewhere. 
\cite{nousek97}$^,$ \cite{weisskopf95}$^,$ 
\cite{zombeck96}$^,$ \cite{markert94}$^,$ \cite{brinkman87}$^,$ \cite{virani2000}

\subsection{High Resolution Camera (HRC)} 
\label{HRC}

The High Resolution Camera, \textit{HRC}, is a microchannel plate
(MCP) instrument. It is comprised of two detector elements, a $\sim$ 
100 mm square optimized for imaging (HRC-I) and a $\sim$ 20 x 300 mm 
rectangular device optimized for the Low Energy Transmission Grating 
(LETG) Spectrometer readout (HRC-S).

The HRC has the highest spatial resolution imaging on \textit{Chandra} --
$\leq$ 0.5
arcsec (FWHM) -- matching the High Resolution Mirror Assembly (HRMA)
point spread function most closely. The HRC energy range extends to
low energies, where the HRMA effective area is the greatest. The HRC-I has
a large field of view (31 arcmin on a side) and is useful for imaging
extended objects such as galaxies, supernova remnants, and clusters of
galaxies as well as resolving sources in a crowded field. The HRC has
good time resolution (16 $\mu$sec), valuable for the analysis of bursts,
pulsars, and other time-variable phenomena but has limited energy
discrimination, \textit{E/$\Delta$E} $\sim$ 1 ($<$ 1 keV). The HRC-S is 
used primarily for readout of the low-energy grating, LETG, for which 
its large format with many pixels gives high spectral resolution ($>$ 
1000, 40-60 $\AA$) and wide spectral coverage (3 - 160 $\AA$).

\subsection{Advanced CCD Imaging Spectrometer (ACIS)} 
\label{ACIS}

\textit{ACIS} is the Advanced CCD Imaging Spectrometer. It is
comprised of two arrays of CCDs, one optimized for imaging wide
fields (ACIS-I; a 2x2 chip array with 16 arcmin on a side), 
the other optimized for grating 
spectroscopy and for imaging narrow fields (ACIS-S; a 1x6 chip array;
8 arcmin x 48 arcmin).
Each array is shaped to follow the relevant focal surface. In 
conjunction with the HRMA, the ACIS imaging array provides
simultaneous
time-resolved imaging and spectroscopy in the energy range $\sim$ 
0.5 - 10.0 keV. When used in conjunction with the High Energy
Transmission Gratings (\textit{HETG}), the ACIS spectroscopic array
acquires high resolution (up to \textit{E/$\Delta$E} = 1000)
spectra of point sources\cite{proposers}. The CCDs had an intrinsic 
energy resolution (\textit{E/$\Delta$E}) which varied from $\sim$5 to 
$\sim$50 across the energy range. However, due to radiation damage
early in the mission, this has been significantly
compromised.\cite{gyp2000}$^,$\cite{virani2001} Nevertheless,
approximately 90\% of all \textit{Chandra} science observation time available
during the first four years of the mission has utilized the ACIS instrument.

ACIS employs two varieties of CCD chips. Eight of the 10 chips are
``front-side'' (or FI) illuminated. That is, the front-side gate
structures are
facing the incident X-ray beam from the HRMA. Two of the 10 chips (S1
and S3) have had treatments applied to the back-sides of the chips,
removing the insensitive, undepleted, bulk silicon material and
leaving only the photo-sensitive depletion region exposed. These
``back-side'', or BI chips, are deployed with the back side facing the
HRMA. BI chips have a substantial improvement in low-energy quantum
efficiency as compared to the FI chips because no X-rays are lost to
the insensitive gate structures but suffer from poorer charge transfer
inefficiency and poorer spectral resolution (prior to launch). 
In addition, early analysis from on-orbit data indicate that
the BI CCDs are more susceptible to ``background flares'', which may
compromise a measurement, than are FI CCDs\cite{plucinsk2000}. These
background flares (\textit{i.e.}, rapid increases in detector background) are
thought to be a consequence of \textit{Chandra's}
radiation environment\cite{virani2000}.

\section{Outline of Data Flow} 
\label{dataflow}

As outlined in the Introduction, there are two data streams that are
monitored for health and safety issues: (1) data received from the
solid state recorder, and (2) ``live'' or ``real-time'' data received
directly from the CXO via the DSN. In this section, we will briefly
outline the data flow associated with the second stream. The former is
presented elsewhere in these proceedings.\cite{spitzbart}

   \begin{figure}
   \begin{center}
   \begin{tabular}{c}
   \psfig{figure=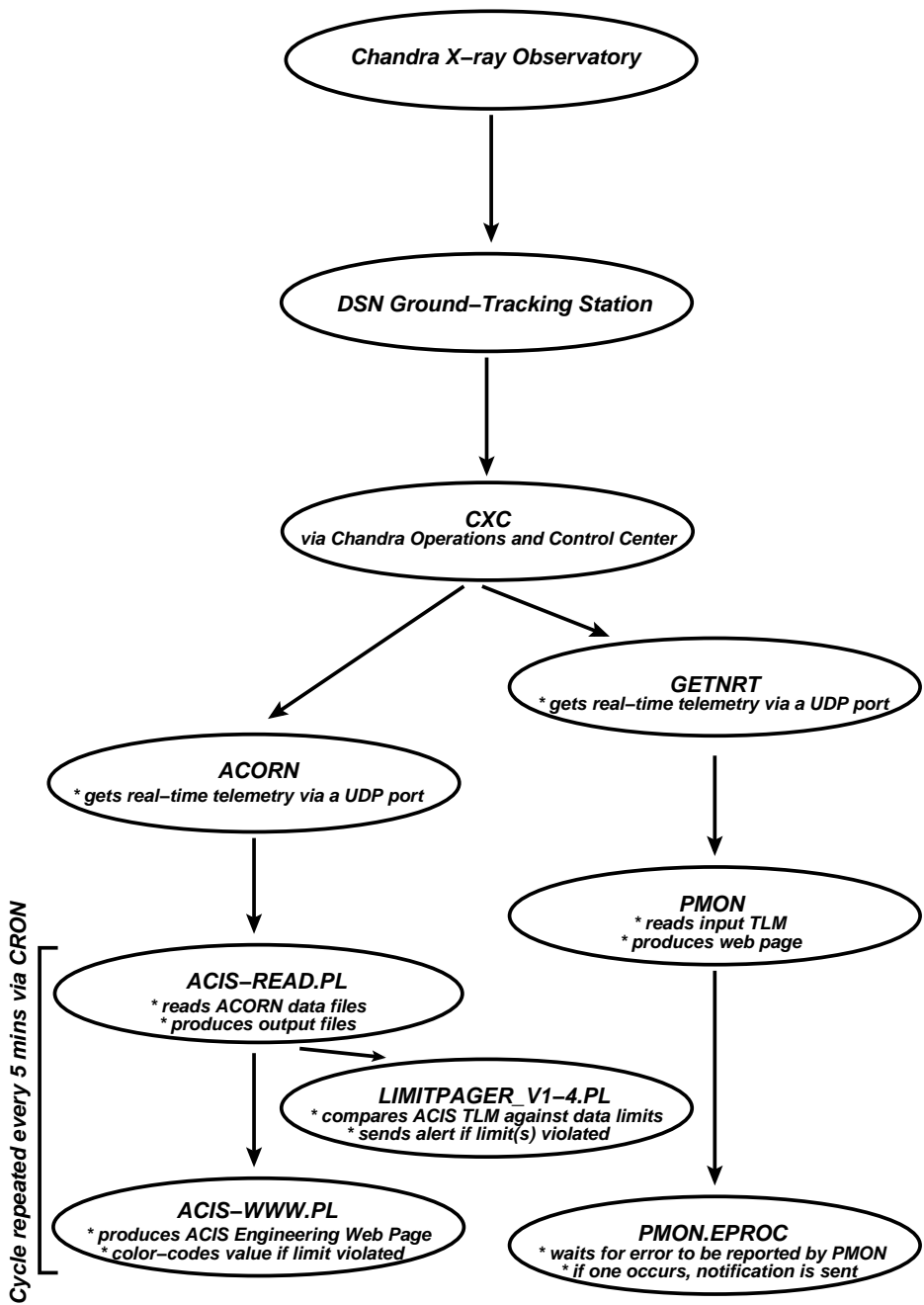,height=20cm} 
   \end{tabular}
   \end{center}
   \caption[example] 
   { \label{RTdataflow}	Real-time Data Flow: The pathway by which
   real-time telemetry from the spacecraft results in ACIS monitoring web pages.} 
   \end{figure} 

Once a communication link has been established with a ground-tracking
station, real-time telemetry begins flowing immediately to the 
CXO Operations and Control Center 
in Cambridge, MA. The
relevant ACIS house-keeping telemetry must be extracted from the raw
telemetry before any health and safety monitoring may be performed.
To accomplish this task, a C++ software package called ACORN (A
Comprehensive object-ORiented Necessity) was written by programmers at
the CXC. 

In our implementation, the real-time raw telemetry stream is fed to
ACORN via a User Datagram Protocol (UDP) port. The raw 
\textit{Chandra} telemetry consists of
approximately 11,000 mnemonic string identifiers (MSIDs). Every
spacecraft and instrument sensor, thermistor, meter, \textit{etc.} are
identified and tracked via a unique MSID. ACORN decodes the real-time
telemetry stream and writes MSIDs, values, and spacecraft times to
either a tab-delimited file or standard output. For our purpose,
ACORN-decoded data is written to a tab-delimited file so that it may
be analyzed. This entire process of real-time raw spacecraft telemetry
to output files that can be processed, is depicted in Figure 1.

\section{Determining ACIS Health and Safety Limits} 
\label{ACIS1}

In order to monitor the health and safety status of the ACIS
instrument, a set of MSIDs as well as a set of limits must first be
established. In total, there are 65 ACIS MSIDs that are regularly 
sent by the instrument to the observatory for incorporation into the
telemetry stream that constitute the ACIS \textit{hardware} ``house-keeping
telemetry''. These MSIDs consist of components 
like the camera-body temperature to the ACIS focal plane temperature
(please see Figure 2). By and large, these are the principle
electronic (\textit{i.e.}, voltages and currents) and temperature MSIDs
reported by ACIS.

Prior to launch, as well as during the first 2 years of the
mission, there were several sets of ``limits'' used by various teams 
within the CXC to ostensibly suit their own particular purpose. The
original health and safety limits (\textit{i.e.}, pre-launch) were
established based on the qualification tests conducted during the
instrumental and observatory thermal-vacuum tests. Therefore,
the first task was to update and streamline the use of various 
limit sets to one definitive set. Two years into the mission, after sufficient 
experience had been gained with instrument response to on-orbit use, 
various members of the CXC convened to identify a robust set of 
limits to check against the real-time ACIS telemetry feed.

   \begin{figure}
   \begin{center}
   \begin{tabular}{c}
   \psfig{figure=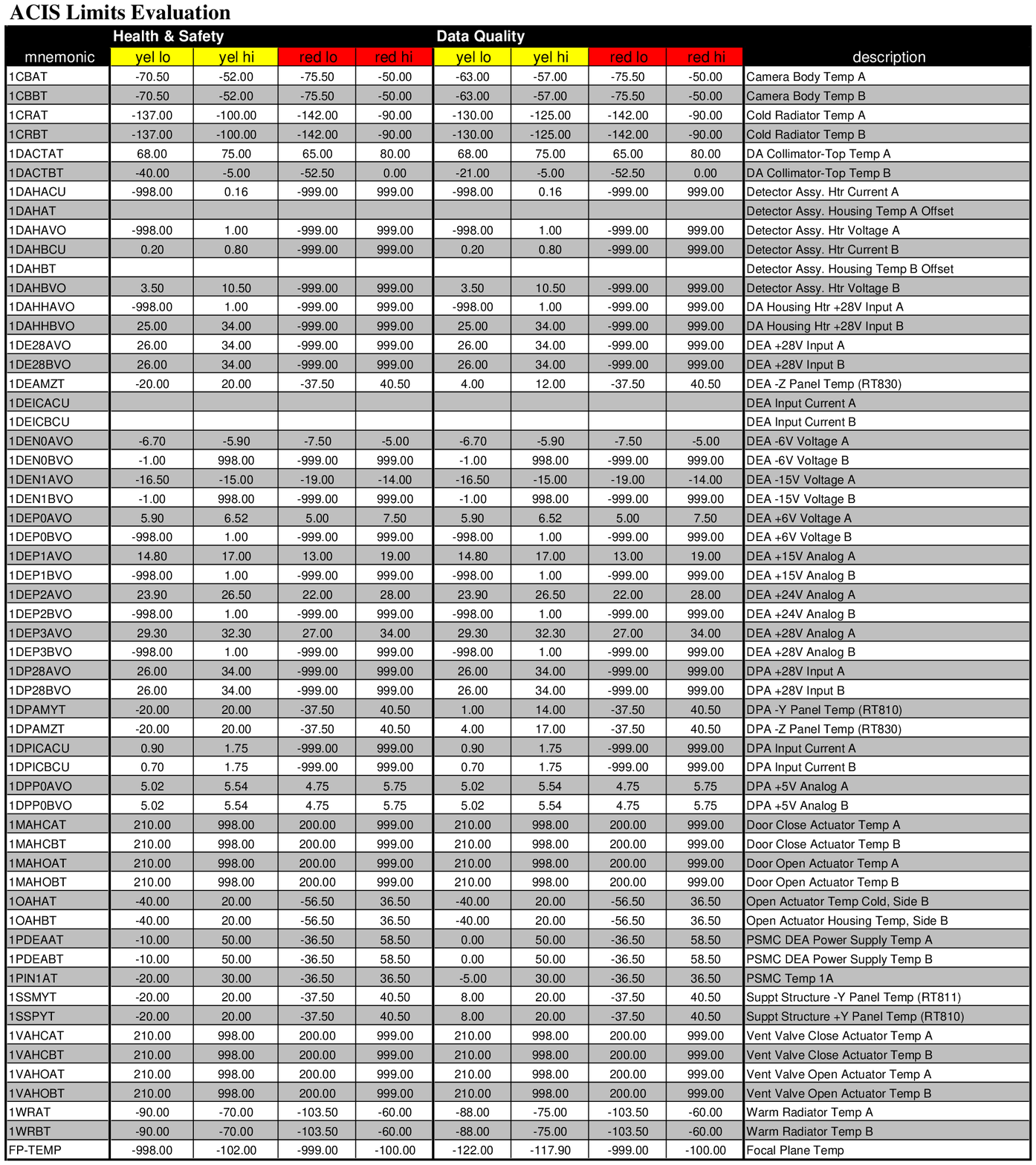,height=20cm} 
   \end{tabular}
   \end{center}
   \caption[example] 
   { \label{limits}	ACIS Health and Safety MSIDs: This listing is
   a compilation of all the 65 ACIS MSIDs that are monitored for health
   and safety reasons. A description and the corresponding ``red'' and
   ``yellow'' limits of each MSID are presented.} 
   \end{figure} 

It was quickly decided that one set of limits would not be desirable
since there are principally two concerns that would govern what the
limits should be: (1) health and safety concerns, and (2) data quality
concerns. Clearly, the former is more important than the latter since
the former is used to monitor and maintain vital mechanical operation
whereas the latter serve to indicate possible reduction in data
quality. Moreover, the philosophy of the the health and safety limits
is such that should a ``red'' limit violation occur, the CXC response
will be immediate. Should a ``yellow'' limit occur, further analysis
may be performed before any response would be initiated.
With this operating principle, two sets of limits were
created for each MSID, with each set containing \textit{caution} 
(color-coded yellow) low and high values and \textit{warning}
(color-coded red) low and high values that would be applicable to
health and safety concerns as well as  concerns that would affect
data quality. 

Nearly 300 days worth of data from the second year of the mission were
analyzed to arrive at the ACIS health and safety as well as data 
quality ``yellow'' and ``red'' limits. Limits from the MIT/ACIS
instrument team were compared against this 300-day baseline and
modified on a case-by-case basis. For some MSIDs, the concept of 
either a low or high ``yellow'' and/or ``red'' limit has no meaning and so
these MSIDs were assigned a nominal value of ``999'' to indicate such
a condition. These limits are presented in Figure 2.
Once this definitive set of limits was determined, they
were implemented by every group within the CXC that monitor the
telemetry from the spacecraft. One caveat that should be noted is that
these MSIDs label telemetered sensor readouts such as the camera-body
temperatures and the various voltages and currents generated by the
ACIS power supply (see figure 2). The ACIS focal-plane temperature,
FP-TEMP, is not included with the MSID values sent to ACORN but is
extracted from the Digital Electronics Assembly (DEA) house-keeping 
packets reported by the ACIS flight software. It is unavailable when 
the flight software is halted or when the DEA house-keeping is disabled.

\section{Real-Time Monitoring of the ACIS Instrument} 
\label{ACIS2}

There are two different ways in which the health and safety of the
ACIS instrument is monitored. The first is by monitoring and
limit-checking the 65 ACIS house-keeping MSIDs discussed in the
preceding section. The second method is by monitoring the ACIS science data
packets produced by the ACIS flight software system. The former is
discussed in Section 5.1 while the latter is discussed in Section 5.2.
The principle difference between these two methods is that the former
monitors telemetry reported by the \textit{hardware} components on ACIS,
whereas the latter monitors the \textit{science} telemetry reported by the ACIS
flight \textit{software}.

\subsection{Monitoring the ACIS House-keeping Telemetry}

With the ACIS health and safety limits as well as access
to real-time telemetry to perform limit-checking established, we now
outline the process by which the ACIS telemetry is monitored and key
ACIS personnel notified in case of an anomaly.

As stated in Section 3, ACORN produces tab-delimited ASCII files for a
set of ACIS MSIDs. ACORN constantly monitors a given UDP port for
spacecraft telemetry. When the telemetry is flowing, \textit{i.e.}
when the OCC has established a communication contact with the CXO, data
from these MSIDs are appended to various ASCII files. When
the OCC is not in contact with the CXO, these ASCII files are not
modified and contain data from the last communication pass. 

Three PERL scripts were written that facilitate the monitoring process.
One PERL script (\texttt{acis-read.pl}) reads the ACIS ACORN log 
files and calculates an 
average and sigma (based on 10 samples of data) for each MSID that is
followed. These values and timestamps are then written to another
tab-delimited ASCII file. The second PERL script
(\texttt{acis-www.pl}) reads these log files
and produces an HTML file that displays this information for team
members to inspect while at work, home, or on travel 
(see http://cxc.harvard.edu/acis). The HTML headers instruct the
user's browser to ask for a new version of the page once every 150
seconds. This second PERL script also incorporates the instrument
``red'' and ``yellow'' limits discussed in the preceding section. If a
given MSID deviates from its nominal value such that it falls
within the ``red'' or ``yellow'' regime, the value for that MSID is
either color-coded ``red'' or ``yellow'' on the web page. A third PERL script 
(\texttt{limitpager\_v1-4.pl}) also reads the ASCII log files produced
by \texttt{acis-read.pl}. It compares the average value for each MSID
reported against its database of ``red'' and ``yellow'' limits. If a
MSID breaches its nominal value such that it falls within the ``red''
or ``yellow'' value span, a pager alert (for ``red limit'' violations) or an
email alert (for ``yellow limit'' violations) is issued by the
script to key ACIS personnel so that they are immediately notified and
so that corrective actions may be taken if required. To reduce the
probability of false alerts due to either spurious data or a ``noisy''
communication contact, the data violation must persist for at least 3
samples before an alert is dispatched.

These three PERL scripts are executed every 5 minutes via a CRON task
maintained by a general group UNIX account. This way the
thorny issue of permission and read/write priviledges is
eliminated. When the OCC is in communication with the CXO, the web
page produced by \texttt{acis-www.pl} is updated with fresh
values. When the OCC is not in communication with the CXO,
\texttt{acis-www.pl} reproduces the last values recorded from the
previous communication contact. Since the beginning and end timestamp
for each MSID value that went into calculating an average and sigma
for that MSID is also presented on the web page, the reader is able to
determine whether the displayed value is representative of the current
instrument status or whether it is an ``old'' value from the preceding
pass.

This web page, the ACIS Engineering Data Web Page, is a key element in
monitoring the health and safety status of the ACIS instrument, and is
the primary means by which ACIS anomalies may be identified. Since
this web page is world-accessible, it allows key personnel of the
CXC to monitor the instrument status whether they are at their office,
at home, or even on travel as long as they have internet access. In
Section 5.1.1, we provide examples of what was discussed in this section.

\subsubsection{Examples} 

In determining the health and safety limits, it was expected that
certain MSIDs would violate their nominal operating range for
certain well-known cases. For instance, it is expected that the cold
radiators on sides A and B (MSID: \texttt{1CRAT} and \texttt{1CRBT}, respectively) would
enter its ``yellow regime'' during perigee transit at certain times of
the year as the instrument tends to warm-up during this point in 
the \textit{Chandra} orbit. Since
the ``red'' and ``yellow'' limits are designed to bracket this MSID's
nominal operating temperature range, we expect a violation to occur at
perigee transit. Please see Figure 3 for an example of what this alert
looks like. Nevertheless, 
since this is a known condition and since
science operation is suspended when the CXO enters the Van Allen
radiation belts, the email alerts that would be spawned by 
\texttt{limitpager\_v1-4.pl} due to ``yellow'' violations by \texttt{1CRAT} and
\texttt{1CBAT} are now suppressed when \textit{Chandra} is
entering, at, or exiting from perigee transit. 

   \begin{figure}
   \begin{center}
   \begin{tabular}{c}
   \psfig{figure=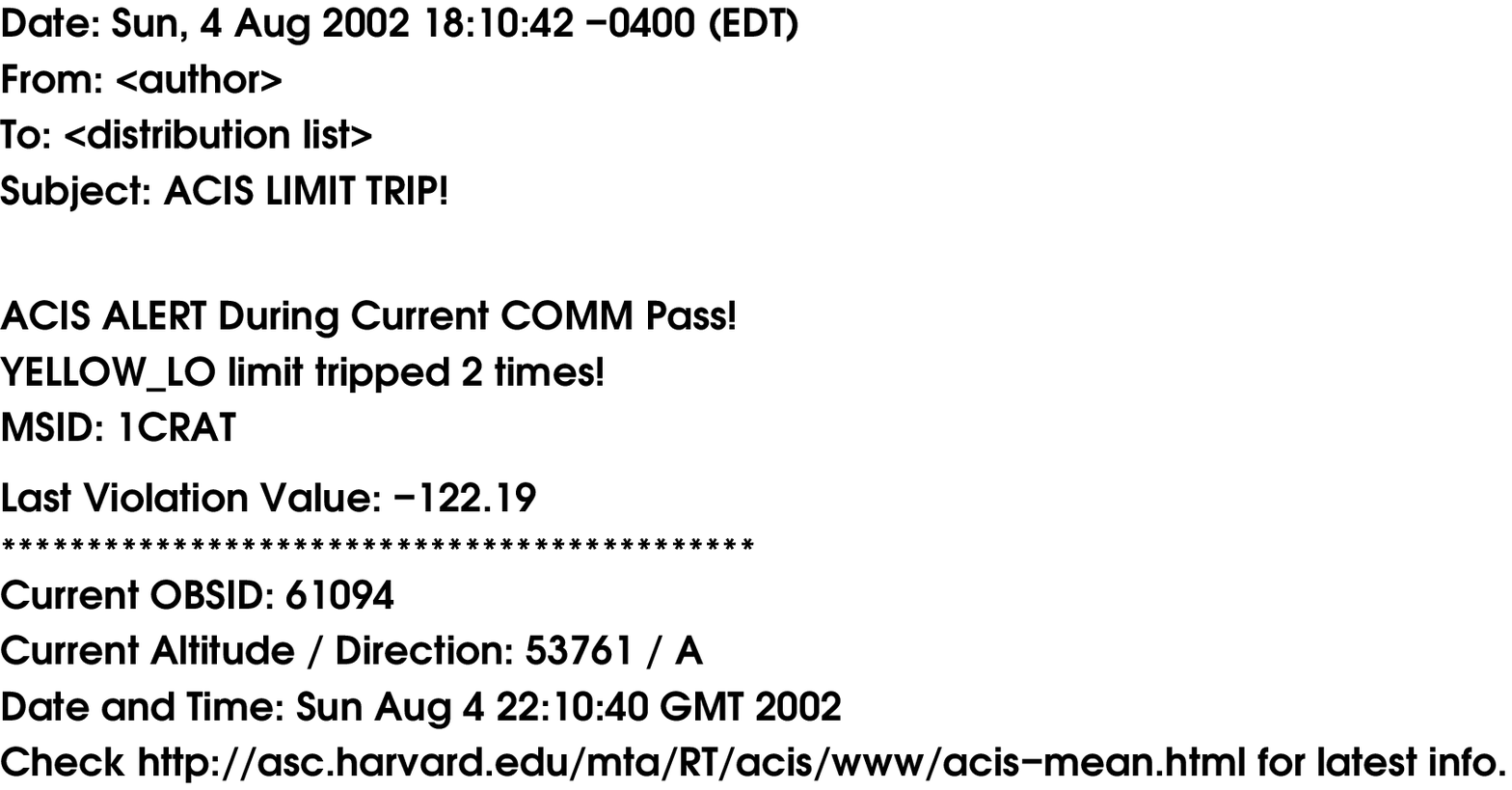,height=7cm} 
   \end{tabular}
   \end{center}
   \caption[example] 
   { \label{alert2}	Email Alert Example: An example of what an
   email alert looks like when the MSID \texttt{1CRAT} breaches its ``yellow limit''.} 
   \end{figure} 

Secondly, as was mentioned in Section 2.2, the spectral resolution of
the front-side CCDs was found to have been significantly reduced early
in the mission. It is believed that the primary mechanism for this
damage is the result of 100-200 keV protons scattering off \textit{Chandra}'s
mirrors and impacting the CCDs while ACIS was in the focal plane during perigee
transit very early in the 
mission.\cite{gyp2000}$^,$\cite{sodell2001}$^,$\cite{kolo2001}
To prevent further damage to the ACIS CCDs,
a prevention scheme was implemented by the Science Operations Team of
the CXC\cite{rac}. One aspect of this scheme has
been to limit the orbital proton fluence and flux as measured by the proton
monitors on-board the ACE and GOES-8 satellites. One way in which the low energy proton
populations can become highly elevated occurs when the Sun unleashes a
coronal mass ejection (CME)\cite{kona}. Once the orbital
fluence has reached its threshold value\cite{rac}, or if one of
the EPHIN channels exceeds its threshold value\cite{rac},
\textit{Chandra} science may be suspended by ground intervention (in
the case of the  former) or autonomously by the spacecraft (in the case of the
latter). When this occurs, the ACIS front-end processors are powered
down and the Digital Processing Assembly's (DPA) input currents decrease such
that a ``yellow'' violation occurs for \texttt{1DPICACU} and \texttt{1DPICBCU}. The
email alerts generated by this violation serve as a confirmation that
\textit{Chandra} has indeed suspended science operations because of
such an event. Figure 4 illustrates what this email alert looks
like when \texttt{1DPICACU} has violated its ``yellow''
limit. Since the ACIS Engineering Data web page makes extensive use of
color and includes various images, an example is not included in this
paper (although the URL is presented in Figures 3 and 4).

   \begin{figure}
   \begin{center}
   \begin{tabular}{c}
   \psfig{figure=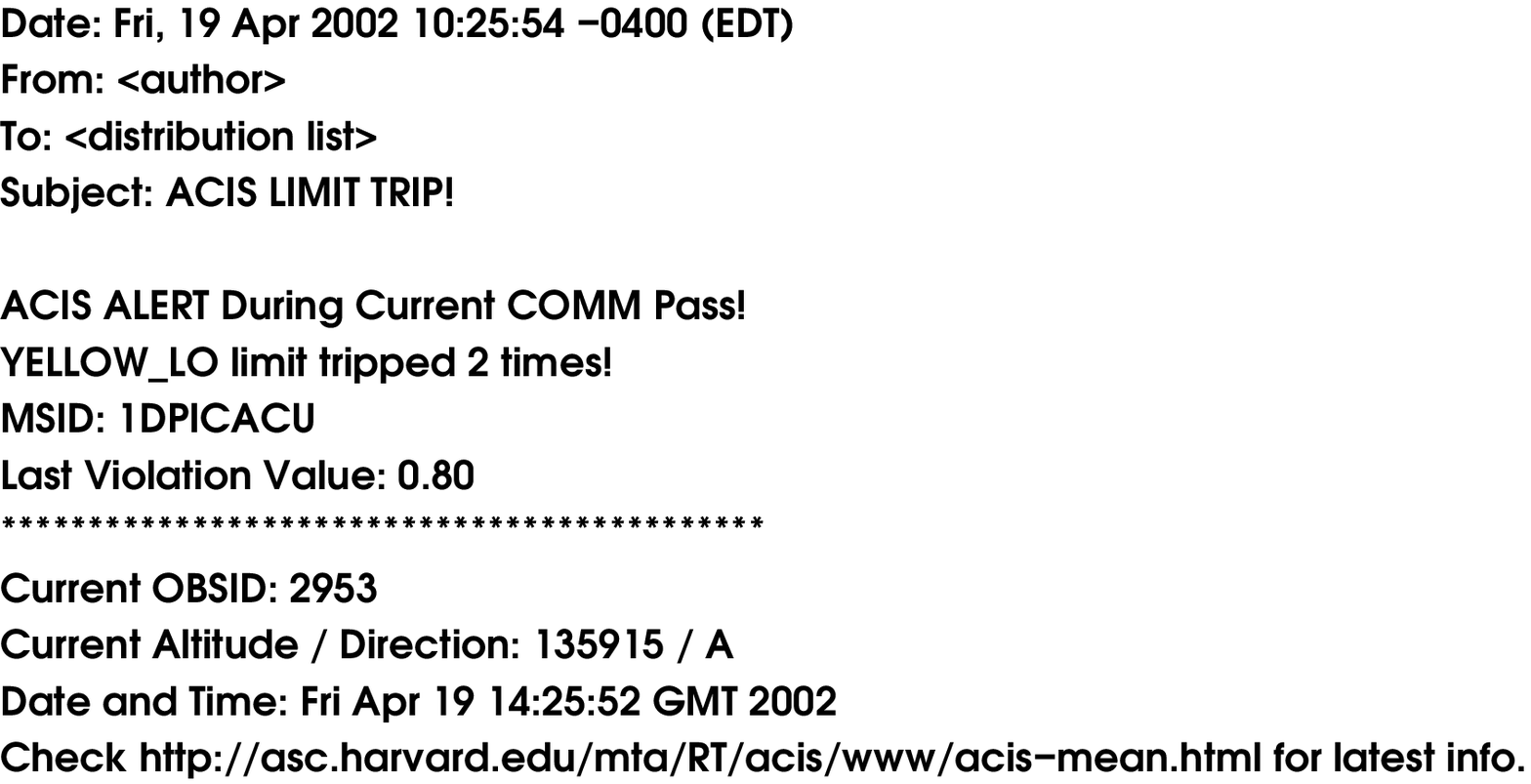,height=7cm} 
   \end{tabular}
   \end{center}
   \caption[example] 
   { \label{alert1}	Email Alert Example: An example of what an
   email alert looks like when the MSID \texttt{1DPICACU} breaches its ``yellow limit''.} 
   \end{figure} 

\subsubsection{Peripheral Issues}

Lastly, there are some peripheral issues that need to be addressed in
order to demonstrate the robustness of the health and safety
monitoring. The first issue is one of redundancy. If the machine that
hosts the CRON task were to shut down either because of an electrical power
failure or due to problems local to that host, the scripts would not
run, the web page would not be produced, and the limit-checking would
not be performed. To solve this problem, we run these PERL scripts and
produce the ACIS Engineering Data web page on two different hosts so
that we are not prone to a single-point failure. One host, which
produces and performs the ``primary'' web page and limit-checking, is
part of a ``mission-critical'' suite of SUN work-stations. Should
this machine go down for any reason, at any time of the day, the Smithsonian
Astrophysical Observatory's computer system group is tasked with the
responsibility of solving whatever problem may afflict the host as
soon as possible. The second host, the ``back-up'', is run on an ACIS group
SUN work-station. Secondly, two additional PERL scripts have been
written that deletes old, obsolete ACORN files as well as ensuring
that the ACORN process is active on the CPU. If the ACORN process is
not found on the CPU, another process is launched. Both of these PERL
scripts are also a part of the CRON task that runs on both hosts. Since
these have been the primary ways in which the web page may not have been
always up-to-date thus far in the mission, we believe we have now
developed a scheme of PERL scripts that is robust to a single-point
failure and is one that can be relied upon to monitor the health and
safety of the ACIS at all times.

   \begin{figure}[htp]
   \begin{center}
   \begin{tabular}{c}
   \psfig{figure=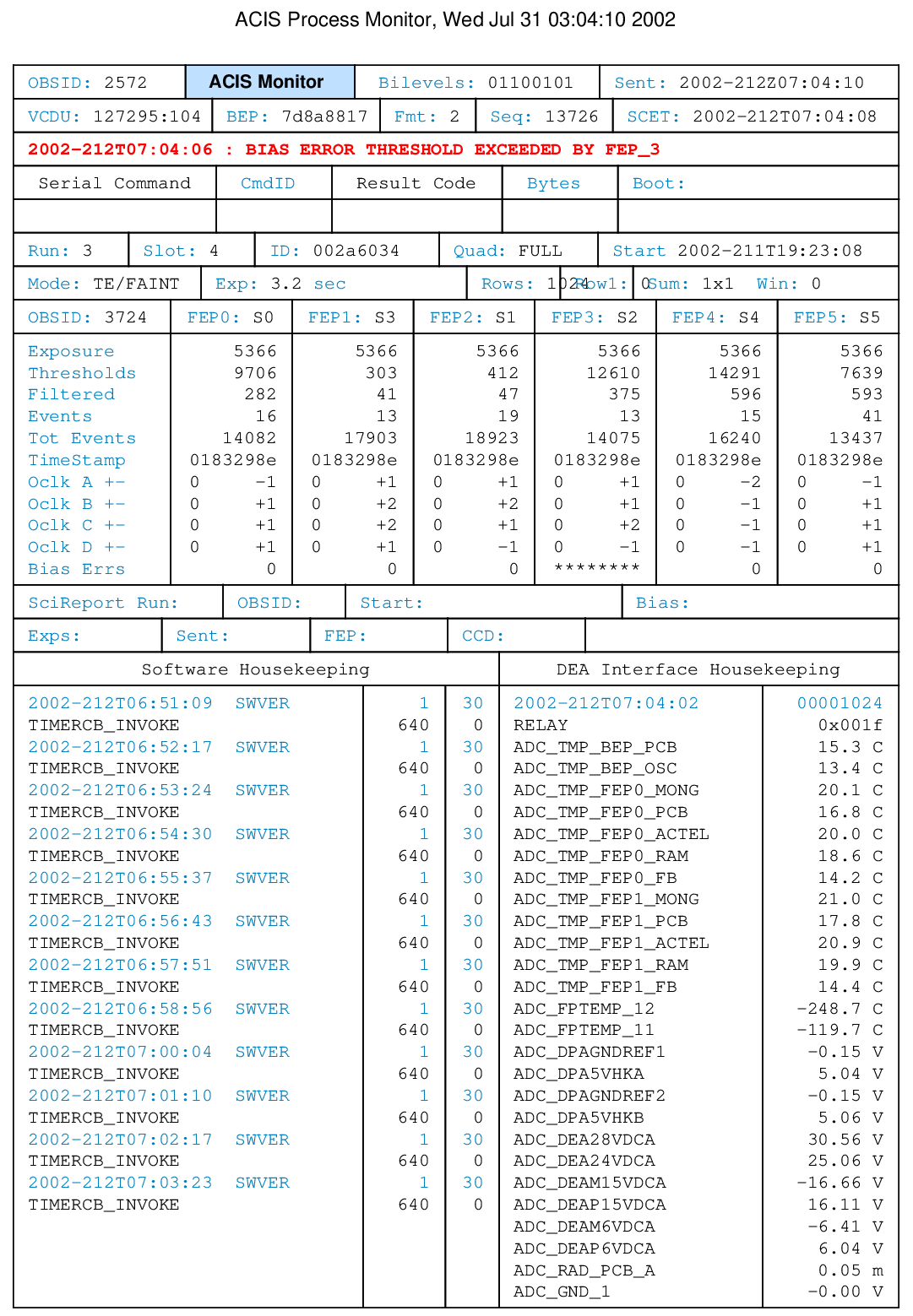,height=20cm} 
   \end{tabular}
   \end{center}
   \caption[example] 
   { \label{pmon}	ACIS PMON Web Page: The monitor receives Chandra telemetry from real-time contacts, extracts ACIS fields, and writes an HTML file.} 
   \end{figure} 

\subsection{Monitoring the ACIS Flight Software Science Telemetry}

The second method by which the health and safety status of the ACIS
instrument is monitored each and every DSN communication pass is via
the ACIS Real-Time Process Monitor (PMON).

The ACIS PMON also receives \textit{Chandra} telemetry during each
real-time contact via a UDP port. A series of shell scripts has been written
that decommutates the raw telemetry, extracts ACIS science fields,
and produces an HTML file
that displays these data. This HTML file presents information that is
relevant to the ACIS science run currently in progress. Information
such as which CCDs are in use, the number of x-ray ``events'' found in
a given exposure frame, the overclock values for each node amplifier
on a given CCD, \textit{etc.}, are displayed. A PMON HTML example is
presented in Figure 5.

When PMON receives fresh telemetry, it writes an HTML file every 30
seconds. Like the ACIS Engineering Data web page, the headers in the 
PMON HTML page instruct the user's browser to ask for a new version 
of the page once every 30 seconds.
However, when not in real-time contact, or in a non-ACIS
telemetry format, some of the displayed values may be very
out-of-date. The user, normally ACIS operations personnel, must
inspect the relevant time fields to determine whether the data are
still useful. As is the case for the scripts used in monitoring the 
ACIS house-keeping electronics and temperatures, a PERL script has
been written that checks to see if the PMON process is active on the
CPU. If it is not, another PMON process is launched. This PERL script
is run every 5 mins via another CRON task.

   \begin{figure}[htp]
   \begin{center}
   \begin{tabular}{c}
   \psfig{figure=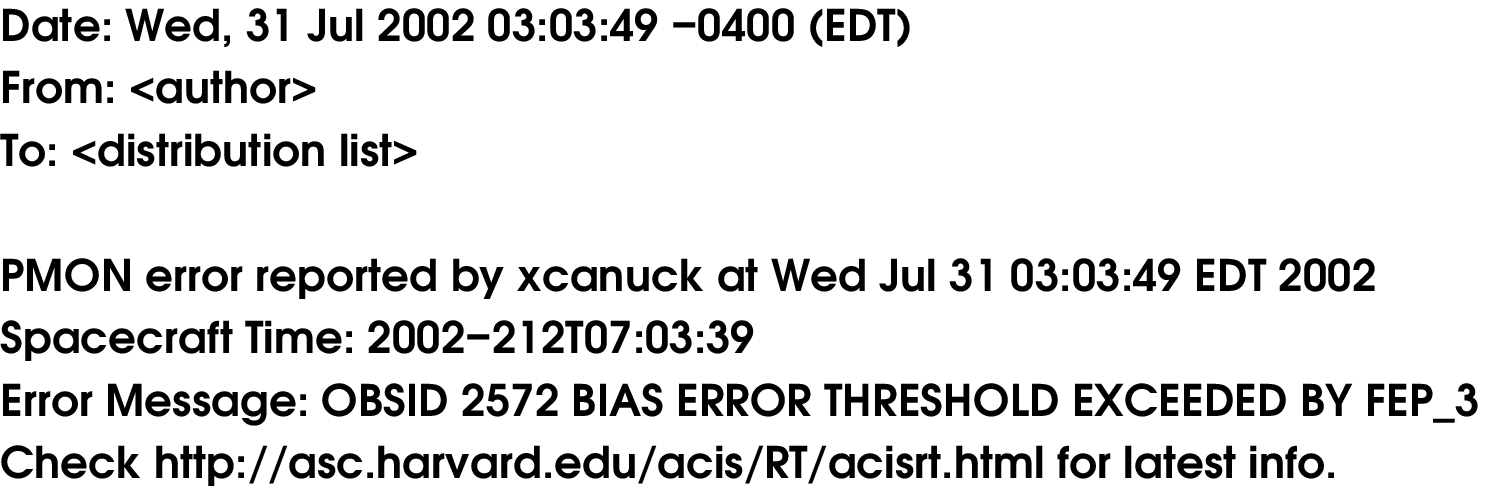,height=4.0cm} 
   \end{tabular}
   \end{center}
   \caption[example] 
   { \label{pmon-error}	PMON Pager Alert: This is an example of a
   recent in-flight anomaly determined via the ACIS PMON and the pager
   alert issued to ACIS personnel to announce its discovery.}
   \end{figure} 

Over the course of three years of successful flight operations, there 
have been three issues that the ACIS Operations team (combining
personnel from the CXC as well as the MIT/IPI team) have had to 
solve with regards to issues of ACIS science operation. One, on a few seemingly
random occassions, the ACIS bias maps have been interleaved with ACIS
science data. The effect of which is to compromise the science data
for that particular CCD and unless corrected, the condition will
persist. Since this anomaly was diagnosed, changes to the way in
which ACIS is commanded via the weekly load have been made such that
we have not seen a repeat of this anomaly, the cause of
which is still unknown. The second issue is one where the bias map
reports a large number of parity errors. The cause of this anomaly is
also not known, however, we have not seen this condition since
November of 1999 and it may have been the result of a single-event
upset. The last issue is one in which a fatal error is
reported by the back-end processor on ACIS. This causes software 
patches to the
ACIS flight software to be lost from memory and requires ground
intervention by the ACIS Operations team to resolve. Since all three 
of these anomalies can be determined via the ACIS PMON, a shell script 
was written that spawns a pager alert (since immediate attention is
required) if one of these anomalies were to occur during a
real-time contact . An example of this type of alert is presented 
in Figure 6 and a comment announcing one of these errors can also be seen
on the PMON HTML page (the third line from the top in Figure 5).

\section{Conclusions}

The ACIS instrument on-board the \textit{Chandra X-ray Observatory} 
has returned some of the most exquisite, dramatic, and detailed images of the 
x-ray universe ever seen. The information returned, and the knowledge
gained from these ACIS observations, has allowed us to explore another
wavelength in rich detail that will undoubtedly propel astronomy
forward in this new millenium. However, the continued success of
\textit{Chandra}, and ACIS in particular,
is dependent upon diligent operation of the spacecraft as well as the prompt 
notification, and hopefully resolution, of any anomalies encountered
in-flight. In this paper, we have presented two distinct methods by
which the telemetry from ACIS during \textit{each and every} 
real-time contact is monitored, and in the event of an anomaly, 
how alerts are quickly dispatched to key ACIS personnel.  As new
anomalies are discovered in-flight, we will endeavour to find ways to
incorporate their monitoring within our current scheme. After three
years of highly successful flight operations, we believe we have
developed a very robust and reliable suite of PERL and HTML scripts that
will help ensure ACIS' continued success.

\acknowledgments     
 
We are grateful to many people, particularly members of the ACIS
Operations team, for their support, encouragement,
fruitful discussions, and suggestions. We thank all of
the engineers, technicians, and scientists who have made the \textit{Chandra}
X-ray Observatory such an incredible success.
The authors acknowledge support for this research from NASA contract 
NAS8-39073.


  \bibliography{report}   

\begin{thebibliography}{10}

\bibitem{sodell98}
S.~L. O'Dell and M.~C. Weisskopf, ``Advanced {X}-ray {A}strophysics {F}acility
  ({AXAF}): {C}alibration {O}verview,'' in {\em X-{R}ay {O}ptics,
  {I}nstruments, and {M}issions},  R.~B. Hoover and A.~B.~W. II, eds., {\em
  Proc. SPIE} {\bf 3444}, pp.~2--18, 1998.

\bibitem{depasq}
J.~M. DePasquale, S.~N. Virani, and P.~P Plucinsky, ``Optimizing the
efficiency of command load inspection for the {A}dvanced {CCD} {I}maging
{S}pectrometer ({ACIS}) on the {C}handra {X}-ray {T}elescope,'' in
{\em {O}bservatory {O}perations to {O}ptimize {S}cientific {R}eturn
III},  P.~J. Quinn, ed., {\em Proc. SPIE} {\bf 4844}, p.~(this volume), 2002.

\bibitem{spitzbart}
B.~D. Spitzbart, S.~J. Wolk, and T.~Isobe, ``{C}handra monitoring,
trending, and response,'' in
{\em {O}bservatory {O}perations to {O}ptimize {S}cientific {R}eturn
III},  P.~J. Quinn, ed., {\em Proc. SPIE} {\bf 4844}, p.~(this volume), 2002.

\bibitem{obsguide}
{\em AXAF Observatory Guide}, Rev. 1.0, Chandra X-ray Science Center,
Cambridge, MA, October 1997.

\bibitem{SIN}
{\em {AXAF} {S}cience {I}nstrument {N}otebook}, Rev. 2.0, Chandra X-ray
  Science Center, Cambridge, MA, April 1995.

\bibitem{nousek97}
J.~A. Nousek, {\em {S}cience {I}nstrument {O}perations
{H}andbook for the {AXAF} {CCD} {I}maging {S}pectrometer}, Version 2.65,
Penn. State Univ., November 1997.

\bibitem{weisskopf95}
M.~C. Weisskopf, S.~L. O'Dell, and R.~F. Elsner, ``Advanced {X}-ray
  {A}strophysics {F}acility - {AXAF} an {O}verview,'' in {\em X-{R}ay and
  {E}xtreme {U}ltraviolet {O}ptics},  R.~B. Hoover and A.~B.~W. II, eds., {\em
  Proc. SPIE} {\bf 2515}, 1995.

\bibitem{zombeck96}
M.~V. Zombeck, ``Advanced {X}-ray {A}strophysics {F}acility {AXAF},'' in {\em
  Proceedings of the International School of Space Science Course on X-Ray
  Astronomy},  {\em Aquila, Italy} {\bf CfA Preprint 403}, 1996.

\bibitem{markert94}
T.~H. Markert, C.~R. Canizares, D.~Dewey, M.~McGuirk, C.~S. Pak, and M.~L.
  Schattenburg, ``{H}igh {E}nergy {T}ransmission {G}rating {S}pectrometer
  ({HETGS}) for {AXAF},'' in {\em {EUV}, {X-R}ay, and {G}amma-{R}ay
  {I}nstrumentation for {A}stronomy {V}},  {\em Proc. SPIE} {\bf 2280}, 1994.

\bibitem{brinkman87}
A.~C. Brinkman, J.~J. van Rooijen, J.~A.~M. Bleeker, J.~H. Dijkstra, J.~Heise,
  P.~A.~J. de~Korte, R.~Mewe, and F.~Paerels, ``{L}ow {E}nergy {X}-ray
  {T}ransmission {G}rating {S}pectrometer for {AXAF},'' {\em Astro. Lett.} {\bf
  26}, p.~73B, 1987.

\bibitem{virani2000}
S.~N. Virani, R.~M\"{u}ller-Mellin, P.~P. Plucinsky, and Y.~M. Butt,
``{T}he {C}handra {X}-ray {O}bservatory's {R}adiation {E}nvironment
and the {AP-8/AE-8} {M}odel,'' in {\em X-{R}ay {O}ptics,
  {I}nstruments, and {M}issions III},  J.~Tr\"{u}mper and
B.~Aschenbach, eds., {\em Proc. SPIE.} {\bf 4012}, p. 669-680, 2000.

\bibitem{proposers}
{\em Chandra {P}roposers' Observatory {G}uide}, Rev. 4.0, Chandra
X-ray Science Center,  Cambridge, MA, December 2001.

\bibitem{gyp2000}
G.~P. Prigozhin, S.~Kessel, M.~Bautz, C.~Grant, B.~LaMarr, R.~Foster,
G.~Ricker, and G.~Garmire, 
``{R}adiation damage in the {C}handra {X}-ray {CCD}s,'' in {\em X-{R}ay {O}ptics,
  {I}nstruments, and {M}issions III},  J.~Tr\"{u}mper and
B.~Aschenbach, eds., {\em Proc. SPIE.} {\bf 4012}, p. 720-730, 2000.

\bibitem{virani2001}
S.~N. Virani, P.~P. Plucinsky, C.~E. Grant, and B.~LaMarr,
``{A}nalysis of {O}n-{O}rbit {ACIS} {S}queegee {M}ode {Da}ta,'' in
{\em X-rays at {S}harp {F}ocus: {C}handra {S}cience {S}ymposium},
E.~M. Schlegel and S.~Dil-Vrtilek, eds., {\em ASP Conference Series}
{\bf 262}, p. 409-418, 2002.

\bibitem{plucinsk2000}
P.~P. Plucinsky and S.~N. Virani, ``Observed {O}n-{O}rbit {B}ackground of the
  {ACIS} {D}etector on the {C}handra {X}-ray {O}bservatory,'' in {\em X-{R}ay
  {O}ptics, {I}nstruments, and {M}issions},  J.~Tr\"{u}mper and B.~Aschenbach,
  eds., {\em Proc. SPIE} {\bf 4012}, p. 681-692, 2000.

\bibitem{sodell2001}
S. L. O'Dell, M. W. Bautz, W. C. Blackwell, Y. M. Butt, R. A. Cameron,
R. F. Elsner, M. S. Gussenhoven, J. J. Kolodziejczak, J. L. Minow, 
R. M. Suggs, D. A. Swartz, A. F. Tennant, S. N. Virani, and K. Warren,
``{R}adiation {E}nvironment of the {C}handra {X}-ray {O}bservatory,'' 
in {\em {X}-{R}ay and {G}amma-{R}ay {I}nstrumentation for {A}stronomy {XI}},
K. A. Flanagan and O. H. W. Segmund, eds., {\em Proc. SPIE} {\bf 4140}, 2001. 

\bibitem{kolo2001}
J. J. Kolodziejczak, R. F. Elsner, R. A. Austin, and S. L. O'Dell,
``{I}on {T}ransmission to the {F}ocal {P}lane of the {C}handra {X}-ray {O}bservatory,'' 
in {\em {X}-{R}ay and {G}amma-{R}ay {I}nstrumentation for {A}stronomy {XI}},
K. A. Flanagan and O. H. W. Segmund, eds., {\em Proc. SPIE} {\bf 4140}, 2001. 

\bibitem{rac}
R.~A. Cameron, D.~C. Morris, S.~N. Virani, S.~J. Wolk,
W.~C. Blackwell, J.~I. Minow, and S.~L. O'Dell,
``{S}helter from the {S}torm: {P}rotecting the {C}handra {X}-ray
{O}bservatory from {R}adiation,'' in {\em ADASS XI Conference
Proceedings}, 2001.

\bibitem{kona}
S.~N. Virani, R.~A. Cameron, P.~P. Plucinsky, R. M\"{u}ller-Mellin, 
and S.~L. O'Dell,
``{M}onitoring the {C}handra {X}-ray {O}bservatory {R}adiation {E}nvironment: 
{C}orrelations between {GOES}-8 and {C}handra/{EPHIN} during DOY
89-106, 2001,'' in {\em Yohkoh 10th Anniversary Science Conference Proceedings}, 2002.


\end{thebibliography}
  \bibliographystyle{spiebib}   
 
  \end{document}